\documentclass{article}

\usepackage{graphicx}
\usepackage{amssymb}
\usepackage{amsmath}

\begin{document}

\title{Complex aspects of gravity}

\author{Mihaela D. Iftime \thanks{Boston, Massachusetts, May 3, 2010}}

\date{}

\maketitle

\begin{abstract}

This paper presents reflections on the validity of a series of mathematical methods and technical assumptions that are encrusted in macrophysics (related to gravitational interaction), that seem to have little or no physical significance. It is interesting to inquire what a change can occur if one removes some of the traditional assumptions.

\end{abstract}

\pagebreak

\section{Determinism or Statistical Dynamics ? }

Whether in general relativity (GR) or quantum theory, nature cannot realistically be described 'from the outside', as if it is seen by a spectator. As imperfect macroscopic observers embedded in the physical world, when we measure a system, we enter into a form of `entanglement' with it. The process of observation is complex and it involves communication and information theory - a description of the phenomena that is intuitively like a 'dialogue' between the observer and the system under consideration.

Determinism is a basic principle in classical physics. If we would have lived in a deterministic world, it would be possible to compute, at least in principle, all past and future behavior, e.g., explain what was before the big-bang, as one might expect from applying the physics laws.

Dirac postulated that classical physics can only predict observables that respect determinism i.e., only gauge invariant functions of the dynamical system variables. \cite{Dirac} The entire constrained system formalism was built on Dirac's approach with the scope of characterizing gauge invariant objects. For example, in the Hamiltonian formalism,  the constraints are characterized by the vanishing Poisson brackets, etc.

Einstein theory of gravitation was confronted with Dirac's principle. Consequently, Einstein formulated the 'hole argument' \cite{Iftime}. Ultimately, Einstein theory   reconciled with determinism: GR observables are the diffeomorphisms (or  spacetime gauge transformations) that displace the spacetime distances between trajectories of particles (or light), in which case, the particles themselves are also displaced under the diffeomorphism action).

\section{Smoothness assumptions.}

The 4-manifold model of space-time has been used for scales, ranging between the radius of elementary particles and the universe.
The gravitational metric field is assumed to be at least $C^2$ (twice  differentiable) to make Einstein equations well-defined. This implies that one could physically determine only $C^3$-atlases. \cite{Hawking}

The smoothness of the metric is not physically important, since one can only measure approximations.

The smooth structure on the 4-manifold however, depends on the atlas, but different (compatible) atlases can give rise to the same smooth structure. While there is a meaningful notion of a $C^k$-atlas, there is no distinct notion of a $C^k$-manifold other than continuous and smooth manifolds. This is because every $C^k$-structure for $k \geq 1$ is uniquely smoothable\cite{Whitney}.
In GR the spacetime coordinates are defined by continuously differentiable (smooth) functions, however this is not a restriction.\cite{Iftime2009}

The smoothness assumption of the null infinity of an asymptotically flat spacetimes  has probably also no real physical significance. Penrose introduced the notion of asymptotic simplicity\cite{Penrose} in order to isolate gravitational systems within the framework of GR, by means of a conformal boundary of the spacetime in the null infinity that has a nice smooth differentiable structure. Examples of on-smooth null infinity are known as polyhomogeneous spacetimes. It is unknown whether the two classes of solutions of the Einstein field equation are large enough to provide a complete description of the gravitational physics of isolated bodies (within GR framework).\cite{Valiente}

The asymptotic simplicity played an instrumental role in providing a suitable framework for the discussion of black hole radiation. Isolating the black hole within an event horizon - the boundary of the future-directed smooth null infinity in a far-field region - allowed the application of linearized approximation of GR and Calculus rules.

\section{Quantum gravity effects?}

Einstein theory treats the universe as it were made up of large-scale continua and draws deterministic conclusions about the world around us. The theory lacks the perception of fine details.

We believe that a quantum theory based on traditional methods of exploring `all possibilities' allowed by the constraints is that the solution space lies in a highly complicated superspace.  As a consequence, the ordinary statical methods cannot be applied in a straightforward manner, and an a priori calculated average over "everything" will generally produce erroneous results.

We have the example of Schr\"{o}dinger equation, which is an excellent theoretic model for the evolution equation to the hydrogen atom, by  providing a complete description of the probabilities should be possible. However, in practice, the solution of Schr\"{o}dinger equation becomes rapidly insoluble for more than one particles.

\section{Dynamics plus thermodynamics?}

Statistical thermodynamics is a ``natural" approach that seeks to explain the microscopic phenomena from a macroscopic perspective. That's why the Second Law of thermodynamics appears in surprisingly so many different forms to explain a variety of complex phenomena.

At the very core of the second law of thermodynamics is a basic property of short-range character of interactions among a very large number of elements, the characteristic of molecular chaos, which is encountered also in other systems in nature. \cite{experimentchaos}

On the other hand, Shannon entropy ( or more exactly, Shannon information-entropy\cite{Shannon}) is a notion of the channel capacity in continuous-time communication, defined as a bound on the maximum amount of error-free units data (or information) that can be transmitted with a specified bandwidth subject to Gaussian noise.
Shannon entropy in the language of probability has been interpreted as a way to describe the uncertainty (or disorder) of the system as the average 'surprisal' for the infinite string of symbols produced by the coding device. The analogy of thermodynamics entropy with Shannon entropy was the doorstep that Boltzmann used in formalizing his physical theory.\cite{Boltzmann}

Boltzmann's second law and his relation between entropy and information has been used as a general rule to many situations, some of them being based on little or inapplicable evidence. For example, black holes thermodynamics\cite{Bekenstein, Hawking}, the principle of maximum entropy\cite{Jaynes}, the principle of extreme physical information\cite{Frieden}, image reconstruction, species
geographic distributions\cite{Schapire}, and complexity of life.

In the case of black holes thermodyamics, the theoretical properties of black holes are deduced from the mere resemblance of the behavior of the surface area of the event horizon of a black hole with the second law of thermodynamics. Ideas were further developed to define the notion of entropy and temperature of a black hole that would emit thermal radiation that could give a characterization of the gravitational sources.\cite{Hawking}

The use of entropy in the Boltzmann's probabilistic model however, requires very restricted assumptions about the preparation of the system and as well on
the nature of the collision mechanism (Markov processes). It can only apply to isolated systems, while most real system in nature are open, and away from thermodynamic equilibrium.\cite{Prigogine}

Boltzmann's  method in physics is a statistical theory, that is phenomenological in nature and concerns with experiences without any hypothesis in dealing with phenomena.

We believe that a complete theory of dynamics must include the phenomenological aspects of evolution of systems,
isolated or not, which can describe the details of the complexity behavior that are inherent in the very structure of the universe and omitted in the traditional dynamics.

\section{Statistical analysis of the likelihood of observations. Maximum Entropy estimator.}

Describing natural phenomena is equivalent to building models and then confronting them with observations.
Gravity posed many difficulties to all varied quantum theoretical developments, which are still inconclusive due to the fact that there are no experiments confirming quantum behavior of space-time.

We shall illustrate with an example how statistical methods can lead to interesting physical results related to complex aspects of gravity in the linear approximation, such as the far field. \footnote{Real gravitational systems that possess a far field linearized limit include: the planetary system or a black hole, for which the closest matter is so far away that the gravitational field in the intermediate region where
the gravitational metric is weak, in the sense that it has only small departures from the flat (Minkowski) spacetime. Of course, for real gravitational systems the far fields doesn't reaches infinity; it is only an idealization.}

The question we pose is related to the likelihood of predicting quantum indeterminacy of the space-time structure, from an overall behavior of the gravitational field at the continuous limit in the far field region.\footnote{
At no stage of our presentation we care about the detail 'nitty-gritty' of the nature of the gravitating bodies with which the gravitation field is associated. Moreover, only the macroscopic state is what really interests us, in particular the most likely observable macrostructure.}

GR describes gravity is in terms of the (gravitational) metric tensor functions, that define local distortions of space and time at a continuous coordinate space-time location. In order to be able to study physics in this 4-dimensional manifold model, we must be able to measure the spatial and temporal separations of neighboring points.

The beauty of investigating far fields is that one can introduces a local inertial system in the neighborhood of a spacetime point-event.
One can imagine a far-field inertial observer that moves along an (arbitrary) temporal world-line and carries with her an orthonormal triad of vectors (whose directions she identifies with the direction of her spatial coordinate axes), assumed to be permanently in the origin of the spatial system, and as time she uses her proper time.

Consider an experiment that performs classical (deterministic) measurements of the weak-field gravitational metric deviations $h(x)$ at a spacetime location $(x)$ using Einstein equations. The quantities $h(x)$ are macroscopic continuous that have a statistical nature, which follows from the assumption that positional accuracy of determining their values is limited to random fluctuations $\delta h$ that occur at at small Planck distances.\cite{Iftime2009}

Our goal here is to determine a statistical measure of likelihood observations of these fluctuations in the form of a macroscopic quantitative based on the classical gravity model. The answer will shed some light on the question of whether or not far-field random fluctuations can in principle be observable.

Note that gravitational waves, as formulated in linearized gravity, have very small amplitudes. In reality this may not be the case. Also, in traditional statistical mechanics one assumes that for a system containing a large number of elements, the fluctuations from the most probable result are expected to be limitingly small. This assumption justifies the use of Shannon entropy and MaxEnt Principle for taking the most probable macrostructure (the one with the most microstates) as representing the so-called equilibrium state of the system.( see \cite{Iftime2009} for a different approach)

Gravitational radiation has not been directly detected, however, it has been indirectly shown to exist\footnote{Hulse-Taylor binary system experiment, 1993}. Other devices to detect gravitational wave motion has been proposed, but they seem inconclusive. (see e.g., Weber experiment - a large, solid bar of metal isolated from outside vibrations \cite{Weber})

Let's start our statistical analysis in the far-fields by looking at a situation we are all familiar: rolling a dice.

 Consider rolling a dice with known, not necessarily fair, {\em a priori} probability $q$.
 Imagine we perform a 'long' experiment to measure a discrete random variable, counting the number of times we observe each face of the die, i.e., the relative frequency distribution  ${f_{i}}$ of each face of the die.

As it is well known, the likelihood of observing the frequencies ${f_{i}}$ conditioned by an a priori model $q$ that actually generated the observations is given by the multinomial likelihood, $ L(f;q): = \displaystyle n!\frac{\prod q^{f}}{\prod f!}$. To understand this equation intuitively, one must notice that the factor $W: =\displaystyle \frac{n!}{\prod f!}$ is the ``multiplicity" of $f$, or the number of combinations that gives rise to the observed (fixed) distribution ${f_{i}}$.
\footnote{Laplace used $W$ instead of the logarithmic form known as Shannon 'entropy' $\displaystyle lim_{n\to\infty} \frac{log W}{n} =-\sum_{i}{p_{i}log p_{i}}$, $\displaystyle lim_{n\to\infty} \frac{log W}{n} =-\sum_{i}{p_{i}log p_{i}}$ } Since with independent observations, probabilities must be multiplied together to recover the joint probability of all measurements, $L$ decreases multiplicative, as $N$ increases. An appropriate statistical measure independent of the number of measurements is the geometric average likelihood $\Upsilon =\displaystyle\sqrt[n]{n!\frac{\prod q^{f}}{\prod f!}}$, a quantity that is $\Upsilon\in[0,1]$, it has value $1$ (certainty) when $p\rightarrow q$, and decreases to $0$, when $p$ diverges from $q$.

An ``entropic'' formulation\footnote{One advantage of using the entropy language is that the KL information can be used with Bayesian strategies that allow a completely explicit procedure for updating the model based on new information.} can be obtained by invoking the Stirling approximation in the limit $n\to \infty$: $log\Upsilon = \displaystyle 1/n(log n! -\sum_{i}{log f_{i}} + \sum_{i}{f_{i}log q_{i}}) $ is given by:

$$\displaystyle log\Upsilon\approx logn - \sum_{i}{p_{i}log f_{i}} + \sum_{i}{p_{i}log q_{i}} = - \sum_{i}{p_{i}log p_{i}} + \sum_{i}{p_{i}log q_{i}} =$$

$$\displaystyle=H(p) - \sum_{i}{p_{i}log q_{i}} = $$

$$- I_{KL}(p\|q) $$
where $\displaystyle H(p)= -\sum_{i}{p_{i}log p_{i}}$ is the Shannon entropy of $p$, and
$\displaystyle I_{KL}(p\|q) =\sum_{i}{p_{i}log\frac{p_{i}}{q_{i}}}$ is the discrete Kullback -Leibler (KL) divergence\cite{KL} of $q$ from $p$ ( where probabilities $p_{i} =f_{i}/n$).

Modeling in continuous time doesn't avoid the complexity of connecting discrete time data to continuous time reality. If we perform a very large number of measurements, we can treat $n$ as a continuous rather than discrete variable.

For infinitely many independent observations, i.e., when $n\rightarrow\infty$, the limiting value of the last (invariant) expression is  $\displaystyle I_{KL} (p\|q)= \int{p(x)log\frac{p(x)}{q(x)}dx} $, the KL-information.

The Shannon and KL information entropies are probabilistic integrals with many statistical interpretations, including likelihood, uncertainty and entropy.

In statistical mechanics, and the generalized Maximum Entropy models, entropy is interpreted as an average uncertainty as the expectation value $\displaystyle H(p)=-\sum_{i}{p_{i}log p_{i}}=\sum_{i}{u_{i} p_{i}} $ of the 'surprisal'  $u_{i} = log \frac{1}{p_{i}}$ to observe the outcome of the experiment.

Here, we define an average uncertainty of the observer in terms of the log-likelihood $- I_{KL} (p\|q)$ as an overall measure of the physical information from the gravitational system that is observable in the far field region. It can be interpreted as the 'extra' amount of information needed to define a detailed microscopic description of the system, that remains un-communicated by a presentation in terms of the macroscopic variables of classical gravitational theory.

The uncertainty interpretation can be intuitively understood from the relation between the KL and Fisher information.(see e.g., \cite{Frieden}). Minimum KL-information, implies maximum Fisher information, which means the probability is steeply sloped about the fluctuations values. So, high average likelihood (of observing infinitely many data with certain fluctuations $\delta x$, if the particular distribution $q(x)$ generated the data) means high predictability of the values the model, i.e., low disorder and determinacy.

\section{Linearity versus nonlinearity.}

In an analytical description of most natural phenomena one encounters strong nonlinearities.
The nonlinearity of Einstein theory distinguishes it from other fundamental physical theories, such as Maxwell equations (of electromagnetism) and Schr\"{o}dinger equation (of quantum mechanics). The study of nonlinear gravitational effects has important applications in cosmology.

However, the nonlinearity of Einstein equations made the task of finding exact solutions difficult. If one might expect to determine approximations of physical reality, however, the majority of gravitational field solutions deal with corrections to behaviors that are simple distortions of linear behavior. Linear approximation also suffices to provide
theoretical explication of most of the experimental GR tests, including: gravitational wave detection, light deflection, perihelion precession, time delay measurements and gravitational lensing.

Moreover, Einstein field equation reduces to Newton's law of gravity by using both linear approximation and the slow-motion approximation. In fact, the gravitational constant  $ G =6.67 \times 10^{-8}  cm^3/ g sec^2$ appearing in the Einstein equations was determined by making these two approximations.

A consequence of the choice of $G$ is the source of the `hierarchy problem' of the fundamental  interactions, since gauge gravitation can be quantized  only  at the  energy level predicted by Planck for Newtonian quantum gravity.\cite{Planck}
(see the braneworld quantum gravity program for details).

We believe that one of the main issues in quantizing gauge models of gravity is that whatever our precision in determining a state of the system, there will always be a characteristic 'small distance', below which one cannot distinguish points on the contracting fiber, and that the exact initial conditions correspond to some idealization.

\section{Mathematical model of physical space and time.}

When one first begins to learn physics is confronted with the concept of a ``field'' i.e., a mode of describing the phenomena of ``interaction at the distance" of particles. Classical fields are ``generalizations'' of the Newtonian instantaneous (contact) interaction of particles.

There are some issues regarding the concept of field. My first concern is related to the underlying nature of the physical space that was retained to be the affine structure (as is it in Newtonian physics). The usefulness of the``free'' vector concept to represent a Newtonian force at its point of application is clear. However, it seems less appropriate to use linear objects to describe relative interactions at the distance. Nevertheless, one learns to manipulate with vectors and tensors, without paying too much attention to the underlying nature of the physical space-time. The main reason to continue to apply linear methods is to make use of a broad variety of already developed techniques, that culminated with the tensorial calculus exploited by Einstein in the construction of GR. My major concern is when one tries to recover the underlying qualitative relationship from an unsuitable mathematical model.

Secondly, since a change in the position of one particle  must influences other particles, but no one can tell how many particles are in the universe, the very definition of the field to describe interaction at the distance, has a degree of uncertainty. 

Thirdly, a metrical structure is a scalar (inner) product ( e.g., in a extrinsic spacetime manifold) that is a secondary structure on the affine physical space. To recognizes this it means to replace the spacetime metric model with
a``gauge-natural" geometric representation of a gravitational field, given by a global cross-section $\sigma$ of the associated fiber bundle $(F^{*}M/G\stackrel{p}{\longrightarrow} M)$ of $G$-related\footnote{The relevant group of general relativity is $G: = SO(3,1)$, or more precisely the restricted Lorentz group $SO^{+}(3,1)$, which is the set of Lorentz transformations preserving both orientation and the direction of time, which is the identity connected component of $SO(1, 3)$} ) Lorentz group $SO^{+}(3,1)$ coframes on $M$. \cite{Iftime2008}

A coframe at $p\in M$ has a 1-jet $u_{p}=j^{1}\phi $ representation in a local (chart) diffeomorphism  $\phi: U_{p}\to \mathbb{R}^4$ such that $\phi(p)=0$. Two spacetime metric structures are equivalent if and only if the corresponding cross-sections are $Diff(M)$-related.\footnote{The group $Diff (M)$ of global spacetime diffeomorphism acts on the space of the cross-sections $\Gamma (F^{*}M/G)$ as follows: $f^{*}(\sigma): = f^{*}\circ \sigma \circ f^{-1}$, where $f^{*}: F^{*}M\to F^{*}M$ is the induced isomorphism on the bundle of coframes. This implies, that for any $f\in Diff(M)$ there is an isomorphism $(f^{*}, f): F^{*}M\to F^{*}M$ defined by
$f^{*}(j^{1}_{p}\phi) = j^{1}_{f(p)}(\phi\circ f^{-1}\mid _{U_{p}})$ such that $f^{*}(P)=P$.} Gravitational fields modeled by global cross-section of $(F^{*}M/G\to M)$ have also a``local representation'' as family of local cross-sections of $(P\to M)$ related via local gauge transformations. Coframes can be easily used in the process of jet prolongations for $G$-structures to define difference relations among the objects.

Consequently, a linear approach to modeling gravitational interaction seems rather inappropriate. Improvements of the spacetime manifold model requires a re-formulation of the concepts of interaction at the distance, referential system and possibly the underlying nature of physical space-time.

\section{Time.}

It is clear that not all physical measurable quantities are Dirac observables. For  example, `time' is a measurable quantity that is not a Dirac observable. In contrast to measurements of the position at a given time, time itself can not be `predicted' - we can only know 'when' we are.

The world of dynamics, be it classical or quantum, describes a time-symmetrical world.
Yet from our everyday experience, past and future plays different roles. Most obvious objections are thermodynamics (e.g., cooling of a hot cup of coffee) and gravity (all things fall down and not up) - processes that always seem to happen irreversibly. The Second Law of thermodynamics\cite{Boltzmann} teaches us that the entropy of an isolated system never decreases, that explains that fluids don't spontaneously compress. In the case of gravity, GR is the most accepted theory of space and time, but it doesn't
predict space-time distances.

Newtonian time is measured on a linear scale axis. In GR, time and space are unified, but GR is fundamentally constructed using a ''linear'' approach, and so time measurements are still estimations of an absolute ratio between the magnitude of a continuous time variable and a unit magnitude. It seems that the unappropriate use of linear methods to describe nonlinear gravitational phenomena is one source of difficulty in defining an``external time''.
Using a nonlinear approach I developed, one might be able to construct a nonlinear scale for time measurements that has an outside zero origin, and a relative unit $r=\frac{t_{2}- t_{1}}{t_{1}} =const$, the points on the scale forming a cyclic group generated by $r$.\cite{Iftime2010a}

\end{document}